# FRET-Amplified Singlet Oxygen Generation by Nanocomposites Comprising Ternary AgInS$_2$/ZnS Quantum Dots and Molecular Photosensitizers

Tatiana Oskolkova[1], Anna Matiushkina[1,2], Lyubov' Borodina[1], Ekaterina Smirnova[1], Antonina Dadadzhanova[1], Fayza Sewid[1,3], Andrey Veniaminov[1], Ekaterina Moiseeva[4], Anna Orlova[1,*]

[1]International Laboratory "Hybrid Nanostructures for Biomedicine", ITMO University, Saint Petersburg 197101, Russia
[2]Division Biophotonics, Federal Institute for Materials Research and Testing (BAM), Berlin 12489, Germany
[3]Faculty of Science, Mansoura University, Mansoura 35516, Egypt
[4]Center for Photonic Science and Engineering, Skolkovo Institute of Science and Technology, Skolkovo Innovation Center, Moscow 143026, Russia

*Corresponding author a.o.orlova@gmail.com

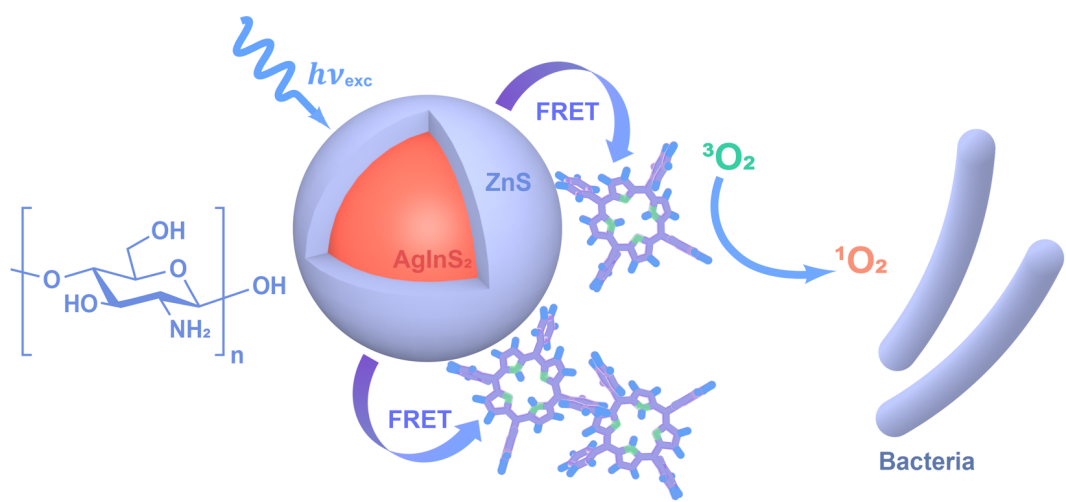

## Abstract

Antibacterial photodynamic therapy (a-PDT) has emerged as a promising non-invasive therapeutic modality that utilizes the combination of a photosensitive agent, molecular oxygen, and excitation light to generate reactive oxygen species (ROS), demonstrating remarkable activity against multidrug-resistant bacterial infections. However, the effective use of conventional photosensitizers is significantly limited by a number of their shortcomings, namely, poor water solubility and low selectivity. Herein, we present a novel biocompatible water-soluble nanocomposite based on hydrophobic tetraphenylporphyrin (TPP) molecules and hydrophilic ternary AgInS2/ZnS quantum dots incorporated into a chitosan matrix as an improved photosensitizer for a-PDT. We demonstrated that TPP molecules could be successfully transferred into chitosan solution while remaining primarily in the form of monomers, which are capable of singlet oxygen generation. We performed a detailed analysis of the Förster resonance energy transfer (FRET) between quantum dots and TPP molecules within the nanocomposite and proposed the mechanism of the singlet oxygen efficiency enhancement via FRET.

# 1 Introduction

The rapid spread of antibiotic resistance among bacterial pathogens has prompted the development of alternative strategies for the diagnosis and treatment of multidrug-resistant bacterial infections.[1,2] Antibacterial photodynamic therapy (a-PDT) is a promising non-invasive treatment modality that utilizes the combination of a light-activated photosensitizer, molecular oxygen, and appropriate excitation sources to generate reactive oxygen species (ROS), such as singlet oxygen (SO), superoxide anion, and hydroxyl radical,[3] which can oxidatively destroy pathogenic bacteria.[4-6] For the past several years, conventional tetrapyrrole-based photosensitizers, which have a high molar extinction coefficient and strong ability to generate ROS, have been widely employed in a-PDT.[7-12] However, the effective use of these therapeutic agents is significantly limited by a number of their shortcomings, namely poor water solubility, rapid photobleaching, and low selectivity of tissue accumulation.[13,14] In this regard, the development of novel biocompatible drug systems for a-PDT, which can overcome the limitations of conventional photosensitizers and significantly increase the efficiency of ROS generation, has become a pressing issue.

The combination of photosensitizers with bioactive natural polymers is a promising approach for improving the properties of conventional therapeutic agents. It has been previously demonstrated that natural biopolymers, in particular, polysaccharide-based polymers exhibit high stability, biodegradability, and non-toxicity, which are attractive qualities for a-PDT applications.[15] Chitosan (CS), also known as poly (1,4,-b-D-glucopyranosamine), is an amino polysaccharide and the second most abundant natural biopolymer.[16] Apart from the biocompatibility[17] and hydrophilicity,[18] CS and its derivatives have been reported to demonstrate antimicrobial activity against several pathogenic bacterial species.[19-22] The positive charge of CS enables it to interact electrostatically with the negatively charged bacterial membrane surface, increasing the probability of cellular uptake.[23] Furthermore, due to the presence of a large number of hydroxyl and amino groups in its structure, CS can be chemically conjugated with photosensitizing agents.[24-27] For example, CS can be used as a matrix for hydrophobic porphyrin molecules, increasing their water solubility,[28,29] as well as it can effectively bind negatively charged nanoparticles.[30,31]

The emergence of nanotechnology in a-PDT has opened up a new area of research on the development of nanocomposites based on photosensitizer molecules and nanostructured materials.[32,33] Quantum dots (QDs) are semiconductor nanocrystals that represent a novel class of highly luminescent nanomaterials.[34] Typically, QDs are characterized by a broad absorption spectral range and high photoluminescence quantum yield along with the great photobleaching resistance and surface modification capability.[35] These features enable QD to act as a donor for the photoexcitation energy transfer to the photosensitizer acceptor through the formation of an additional non-radiative energy relaxation channel that competes with the QD photoluminescence.[36-38] In this case, Förster resonance energy transfer (FRET), which arises from long-range dipole-dipole interactions of the proximal donor and lower energy acceptor, can serve as a dominant relaxation channel.[39] It has been previously reported that the conjugation of QDs with tetrapyrrole-based photosensitizers can expand their excitation range and significantly enhance the efficiency of ROS generation via FRET.[40-44] For example, Sewid et al. have demonstrated that the formed conjugates based on CdSe/ZnS QDs and chlorin e6 (Ce6) molecules exhibit a twofold FRET-mediated enhancement of the efficiency of SO generation as compared to free photosensitizer molecules.[45] In another study, Sangtani et al. have designed a QD-Ce6 nanocomposites decorated with membrane-tethering peptide, and tested it on mammalian cells.[46] It has been demonstrated that excitation of Ce6 in a FRET configuration results in membrane-localized ROS generation and inhibition of cellular proliferation.

However, it should be noted that in the majority of studies conventional cadmium-based QDs have been used for the conjugate formation. Although such QDs represent a class of semiconductor nanocrystals with the most well studied optical and electrochemical properties, the presence of toxic heavy metals in their structure significantly hinders their further clinical applications.[47] Recently, ternary I–III–VI QDs such as $CuInS_2$[48] and $AgInS_2$[49] have gained much attention as a non-toxic alternative to conventional binary II–VI and IV–VI nanocrystals. $AgInS_2$ QDs are promising due to their wide band gap ranging from 1.8 to 2.0 eV[50], which provides tunable emission in the visible and near-infrared regions, in particular, in the 600-800 nm optical window suitable for bioimaging.[51,52] Besides, $AgInS_2$ QDs are also attractive for the FRET-based systems formation, since, due to their broad photoluminescence spectrum, more absorption bands of the different acceptors could possibly overlap with the emission of the donor.[53] Notably, many intriguing features of ternary $AgInS_2$ QDs, as well as the mechanisms behind their photoluminescence,[49,54] remain unclear, leading to an increasing interest towards the photophysical studies of these nanocrystals.

Herein, we present a novel biocompatible CS-passivated nanocomposite based on hydrophobic tetraphenylporphyrin (TPP) molecules and hydrophilic ternary $AgInS_2$/ZnS QDs for a-PDT applications. We investigate the optical properties of the nanocomposites by the means of the steady-state spectroscopy and estimate the fraction of luminescent TPP monomers in the CS matrix. Then, we discuss the results of the time- and spectrally-resolved photoluminescence study of the nanocomposites and analyze the quenching rate of the different radiative transitions in QDs in the presence of TPP molecules. The investigation of the electronic relaxation processes is crucial as they play a key role in the interpretation of the mechanisms underlying the functionality of the novel nanostructures for a-PDT applications. Next, we calculate the efficiency of FRET within the nanocomposite and estimate the enhancement of SO generation by the hybrid structures with QDs compared to the free TPP molecules in CS.

## 2 Results and discussion

### 2.1 Characterization of the CS-AIS-TPP Nanocomposites

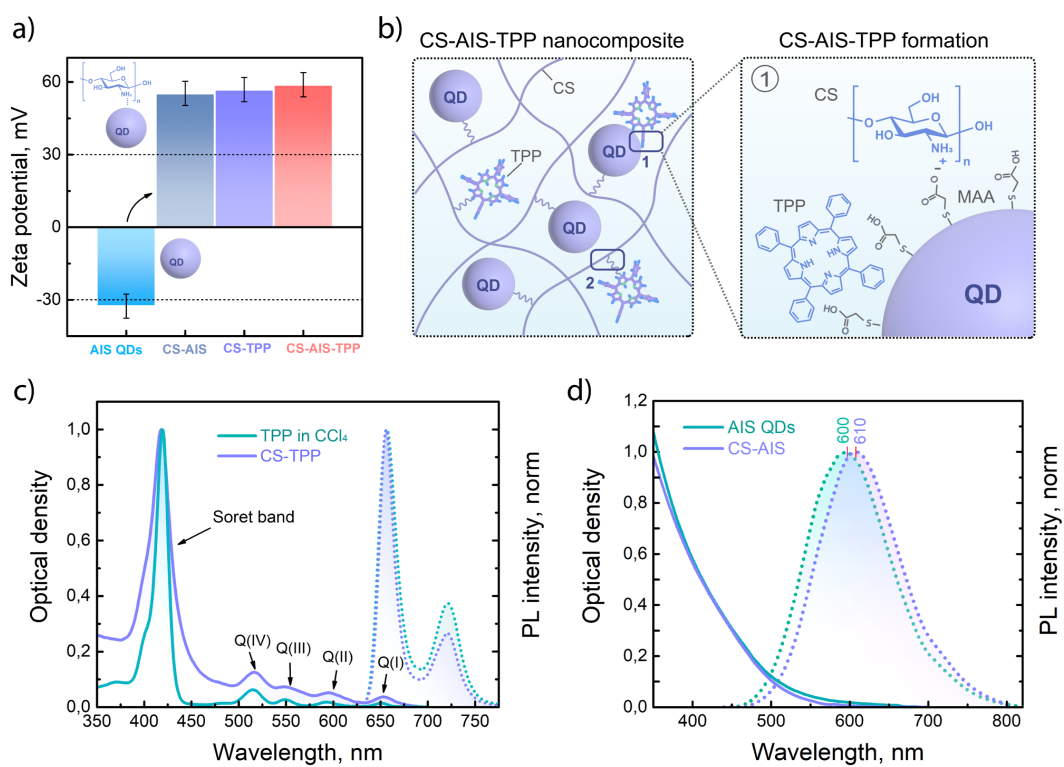

**Figure 1.** (a) Zeta potential of the samples. (b) Schematic illustration of the AIS QDs and TPP molecules distribution within the CS-AIS-TPP nanocomposite. In the (1) configuration the TPP molecules are bound to the AIS QDs surface, while in the (2) configuration TPP molecules are bound only to the CS functional groups. Absorption and photoluminescence spectra of (c) TPP in $CCl_4$ and CS-TPP solution; (d) AIS QDs and CS-AIS solution. The photoluminescence excitation wavelength is 420 nm.

To investigate the size properties of the formed CS-AIS-TPP nanocomposites the well-established Dynamic Light Scattering (DLS) technique has been employed. Using this technique, the average hydrodynamic diameter of the nanocomposites has been determined to be approximately (140 ± 30) nm (Figure S1a). To further gain a better understanding of the morphology of the formed nanocomposites, the Stripe Fluorescence Recovery After Photobleaching (s-FRAP) technique has been utilized (See Section S1 in the ESI for details). By using s-FRAP, we not only obtained additional information about the size characteristics of the nanocomposites but confirmed their ability to exhibit luminescence. As a result of the measurement, a hydrodynamic diameter value of (160 ± 30) nm for the CS-AIS-TPP nanocomposites in a colloidal solution has been obtained. To note, this result is in a close agreement with the findings derived from the DLS studies, which further validates the accuracy of the measurements. Additionally, the absence of any signature of the smaller luminescent components indicates the complete inclusion of AIS QDs and TPP molecules in the CS matrix.

Zeta potential measurements are useful for understanding and predicting interactions between components within a colloidal system and can provide necessary information on the stability of the samples.[55] As shown in Figure 1a, the negative zeta potential values have been obtained for AIS QDs, which can be explained by the presence of negatively charged carboxyl groups of the mercaptoacetic acid (MAA) ligands on their surface. The incorporation of AIS QDs into the CS matrix results in the change of zeta potential value from negative to positive. It should be noted that the average zeta potential of all of the CS-passivated structures exceeds the value of +30 mV, indicating their good colloidal stability.[56] We assume that the distribution of AIS QDs and TPP molecules within the CS matrix can be represented by two possible configurations (Figure 1b). In the first case, the TPP molecules are directed bound to the AIS QDs through the coordination of TPP phenol groups onto zinc atoms of the QD shell.[57] The bond between CS and AIS QDs is formed by the electrostatic interaction between the amino groups of CS and the carboxyl groups of the stabilizing MAA on the QD surface. In the second case, both TPP molecules and AIS QDs are attached only to the CS functional groups without binding to each other. Notably, in this configuration the components of a nanocomposite may still appear to be present at a relatively short distance from each other due to the CS ability to self-assemble.[58] So, we propose that both configurations can coexist in the formed nanocomposites.

Figures 1c and 1d represent the optical properties of TPP molecules and AIS QDs before and after incorporation into the CS matrix, investigated by steady-state UV-VIS and photoluminescence spectroscopy. As can be seen (Figure 1c), hydrophobic TPP molecules dissolved in tetrachloromethane ($CCl_4$) are characterized by a strong absorption band with a maximum at 420 nm, the so-called Soret band, and four weak Q-bands. Except for a slight broadening of the Soret band, which can be related to the appearance of TPP aggregation,[59] the incorporation of TPP into the CS matrix has not led to drastic changes in their absorption spectrum. Moreover, the peak positions of the emission bands of TPP are preserved. As for the AIS QDs (Figure 1d), their absorption represents a broad spectrum without a sharp excitonic band, which is typical for ternary QDs.[49] The photoluminescence spectrum of the QDs has a maximum at 600 nm and undergoes a red shift of 10 nm when transferred to CS.

To evaluate the effect of the TPP concentration on the efficiency of the non-radiative processes within the CS-AIS-TPP nanocomposites and thus their potential performance in a-PDT, we varied the molar ratio of TPP molecules and AIS QDs and investigated the photophysical properties of a series of samples. Figure 2a shows the absorption spectra of CS-AIS-TPP nanocomposites with different $n = C_{TPP}:C_{AIS}$ molar ratios. It can be clearly seen that the nanocomposites retain the spectroscopic properties of both TPP molecules and AIS QDs, and their absorption spectra can be represented as a superposition of the CS-AIS and CS-TPP spectra. As the concentration of TPP increases, the optical density of the characteristic Soret and Q-bands exhibit a regular increase. These observations indicate that the independent CS-embedded components do not undergo significant modifications upon the nanocomposite formation. Furthermore, the progressive photoluminescence quenching of AIS QDs with increasing content of TPP molecules has been observed (Figure 2b). Similar findings can be obtained by analyzing the photoluminescence excitation spectra of the nanocomposites presented in Figure 2c. As can be seen, the reflectance of the TPP emission becomes more pronounced with increasing the amount of TPP molecules, while the contribution of the AIS QDs photoluminescence exhibits a gradual decrease. There are a number of processes that can generally be responsible for the decrease in the photoluminescence intensity of QDs in the presence of molecular quenchers, such as photoinduced electron transfer,[60] the formation of local surface trap states in QDs,[61] and non-radiative energy transfer.[36,37] However, the probability of the first two processes occurring in the formed nanocomposites is rather low, since they require a direct attachment of the molecules to the QD surface. Moreover, the electron transfer in the complexes based on QDs and porphyrin derivatives has not been experimentally confirmed to date. The non-radiative energy transfer occurring through the FRET mechanism can serve as a dominant quenching process in the formed nanocomposites due to several reasons. Firstly, although the FRET can occur at distances up to 10 nm, the direct contact of the donor and acceptor of energy is not required, which is favorable for the formed nanocomposites. Secondly, we satisfy the FRET condition on the finite overlap between the absorption spectrum of the TPP acceptors and the photoluminescence spectrum of AIS QD donors (Figure S2). Finally, a sensitization of the TPP acceptor upon the QD donor excitation at 460 nm (Figure S3) provides good evidence that a donor quenching is caused by FRET and not by any other quenching mechanisms.[36]

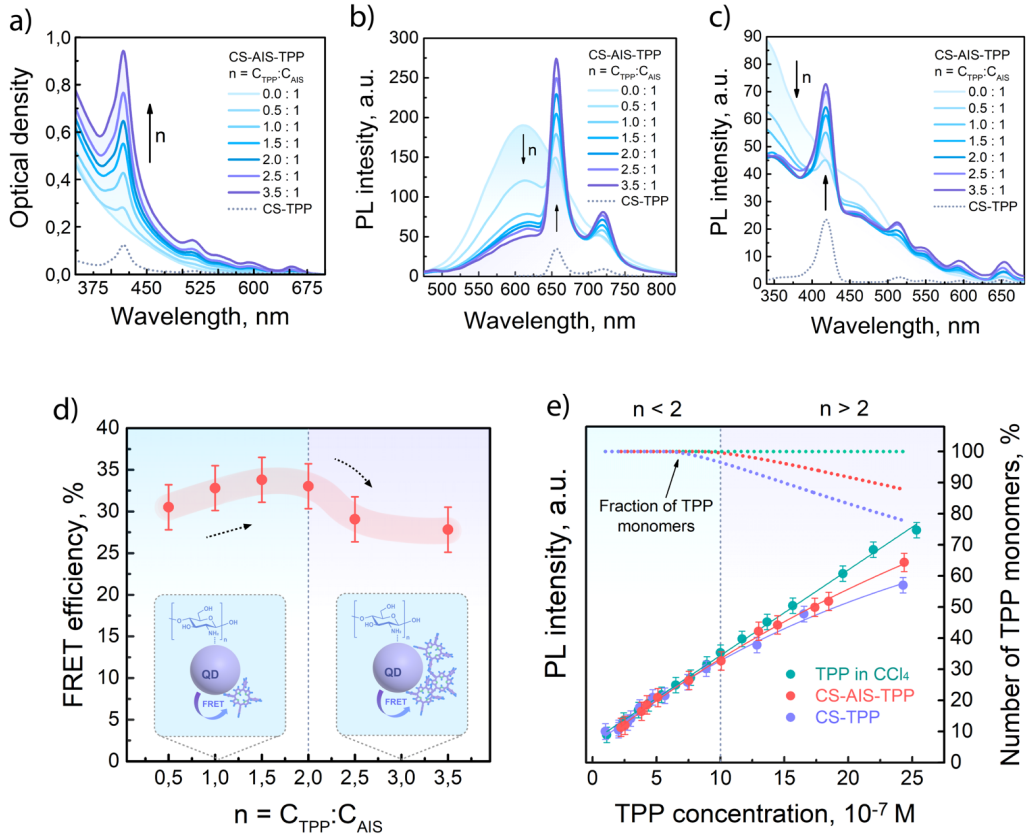

**Figure 2.** (a) Absorption, (b) photoluminescence, and (c) photoluminescence excitation spectra of CS-TPP and CS-AIS-TPP nanocomposites with different n = $C_{TPP}:C_{AIS}$ molar ratios. The photoluminescence excitation and registration wavelengths are 420 and 660 nm, respectively. (d) The dependence of the FRET efficiency between AIS QDs and TPP molecules within the CS-AIS-TPP nanocomposites on the n = $C_{TPP}:C_{AIS}$ molar ratio. (e) The dependence of the intensity of the TPP emission band at 660 nm (solid line) and the number of TPP monomers (dotted line) on the TPP concentration. The photoluminescence excitation wavelength is 600 nm.

## 2.2 Estimation of the FRET Efficiency in the CS-AIS-TPP Nanocomposites

From the above discussion, it is clear that the photoluminescence quenching of AIS QDs in the presence of TPP molecules occurs mainly through a non-radiative energy transfer mechanism known as FRET, which is promoted by a long-range dipole-dipole coupling between a donor and acceptor of energy.[39] The FRET efficiency can be calculated from the relevant spectroscopic properties of the involved donor and acceptor as follows:[62]

$$E_{FRET} = \frac{I_{AD}(\lambda_D^{ex}) \cdot D_A(\lambda_A^{ex})}{I_A(\lambda_A^{ex}) \cdot D_D(\lambda_D^{ex}) \cdot F}, \quad (1)$$

where $I_{AD}$ is the photoluminescence intensity of the sensitized acceptor, $I_A$ is the photoluminescence intensity of the directly excited acceptor; $D_A$ and $D_D$ are the optical densities of the acceptor and donor at the photoluminescence excitation wavelengths, $\lambda_A^{ex}$ and $\lambda_D^{ex}$ are the photoluminescence excitation wavelengths of the acceptor and donor, respectively. $F$ is the photoluminescence quenching efficiency of the donor, given by:

$$F = 1 - \frac{I}{I_0}, \quad (2)$$

where $I$ and $I_0$ are the donor photoluminescence intensities in the presence and absence of the acceptor, respectively. Following this procedure, two different excitation wavelengths have been used to selectively excite the photoluminescence of either the donor QDs or the acceptor TPP within the nanocomposites. The 460 nm excitation wavelength allows direct excitation of the photoluminescence of the AIS QDs due to the presence of a local minimum of TPP absorption at this wavelength. For the direct excitation of the TPP emission, the 600 nm wavelength has been used, given that the absorption of AIS QDs is negligible in this spectral region. It should be noted that the evaluation of the sensitized photoluminescence intensity of TPP molecules upon excitation at 460 nm is somewhat complicated due to the spectral overlap of the AIS QDs and

TPP emission bands. We account for this by subtracting the normalized photoluminescence spectrum of CS-AIS solution from that of CS-AIS-TPP nanocomposites, thus obtaining the emission of individual TPP molecules. The results of this FRET analysis for the CS-AIS-TPP nanocomposites with different n = $C_{TPP}:C_{AIS}$ molar ratios are presented in Figure 2d and Table S1. As can be seen in Figure 2d, the FRET efficiency exhibits a slight increase from 30.5% to about 33.7% for molar ratio values of 0.5 < n < 2.0, which is consistent with the idea that the increase of the number of acceptors leads to the enhancement of the FRET since more energy transfer pathways are available. However, it has been demonstrated that the FRET efficiency further decreases to about 27.8% for 2.0 < n < 3.5. We assume that this phenomenon is attributed to the formation of non-luminescent TPP aggregates[63] as the concentration exceeds the value corresponding to the n = 2. The effect of aggregation on the photophysical properties of TPP molecules and its correlation with the FRET efficiency will be discussed in more detail below (Section 2.3).

The FRET efficiency can also be theoretically evaluated using an alternative approach, following the formula:[62]

$$E_{FRET} = \frac{nR_0^6}{nR_0^6 + R^6}, \qquad (3)$$

where n is the number of acceptors per donor; $R$ is the distance between the donor and acceptor; $R_0$ is the Förster distance, i.e., the distance between the donor and acceptor at which the FRET probability is equal to the probability of the spontaneous deactivation of the excited state of the donor and is given with:

$$R_0^6 = \frac{9000 \cdot \ln 10 \cdot \Phi^2 \cdot QY_{QD}}{128 \cdot \pi^5 \cdot n^4 \cdot N} \cdot I, \qquad (4)$$

where $\Phi^2$ is the dipole orientation factor; $QY_{QD}$ is the quantum yield of the donor in the absence of a quencher; $n$ is the refractive index of the medium; $N$ is the Avogadro number; $I$ is the overlap integral between the donor emission band and the acceptor absorption band estimated as follows:

$$I = \int I_D^N(\nu) \cdot \varepsilon(\nu) \cdot \nu^{-4} \, d\nu, \qquad (5)$$

where $I_D^N(\nu)$ is the normalized photoluminescence spectrum of the donor; $\varepsilon(\nu)$ is the molar extinction coefficient of the acceptor; $\nu$ is the wavenumber.

Assuming that the energy transfer via the FRET mechanism is the dominant non-radiative relaxation process occurring in the CS-AIS-TPP nanocomposites, we can correlate the two approaches for the FRET efficiency estimation and determine the average distance R between the AIS QDs and TPP molecules in a CS matrix. Employing this strategy, we demonstrate that the AIS QDs and TPP molecules are located at an average distance of 9.4 nm within the nanocomposite (Table S2). This value exceeds the average radius of AIS QDs $r_{AIS}$ = 1.35 nm, which confirms our assumption that not all of the TPP molecules are directly attached to the QD surface in a part of the CS-AIS-TPP nanocomposite. This observation suggests the coexistence of both previously discussed configurations of the nanocomposite.

**2.3 Aggregation of the TPP Molecules in the CS-AIS-TPP Nanocomposites**

Porphyrins and their derivatives are known to have a tendency to form non-luminescent supramolecular aggregates in solution.[59] It is noteworthy that the aggregation of porphyrin molecules leads to the loss of their ability to generate SO due to the increase in non-radiative rates, which is usually unfavorable for practical applications.[64] Hence, it is important to estimate the concentration of luminescent TPP monomers in the prepared nanocomposite samples and to elucidate the optimized parameters for the CS-AIS-TPP nanocomposites formation. For this reason, we have analyzed the photoluminescence quenching rate of TPP molecules in a part of the CS-TPP and CS-AIS-TPP nanocomposites as a function of the molecular concentration, using TPP dissolved in an organic solvent as a reference. As shown in Figure 2e, the intensity of the emission band of TPP in $CCl_4$ increases linearly with increasing the concentration, demonstrating that the value of the photoluminescence quantum yield of the TPP molecules remains unchanged. In the case of CS-TPP and CS-AIS-TPP nanocomposites, the deviation from the function linearity can be observed for TPP concentration values above $C_{TPP}$ = $10^{-6}$ M, indicating the occurrence of the aggregation-induced photoluminescence quenching. Based on these results, we have estimated the fraction of the TPP monomers in a part of the CS-TPP and CS-AIS-TPP nanocomposites by dividing the values of their photoluminescence intensities by the reference intensity of TPP molecules dissolved in $CCl_4$. As can be seen in Figure 2e, the TPP molecules of the concentration lower than $C_{TPP}$ = $10^{-6}$ M remain completely in the monomeric form within the

CS-AIS-TPP nanocomposite, while the increase of concentration to $C_{TPP}$ = 2.5 · 10$^{-6}$ M reduces the number of the molecular monomers to about 88%. Of particular note is that the aggregation rate of TPP in the CS-AIS-TPP nanocomposite is lower than that in the CS-TPP solution of the same concentration, where only 75% of molecular monomers are preserved. This observation can be attributed to the proposed configuration of the CS-AIS-TPP nanocomposites, wherein the TPP molecules are directly attached to the surface of the AIS QDs, resulting in limited aggregation. Therefore, the samples of CS-AIS-TPP nanocomposites with a molar ratio of n = $C_{TPP}$:$C_{AIS}$ of less than 2 have been found to be the most suitable for further analysis, since they contain entirely monomeric TPP molecules capable of the efficient SO generation. Furthermore, it is important to highlight that the observed relationship between the fraction of TPP monomers and the molecular concentration is consistent with the previously discussed dependence of the FRET efficiency on the n = $C_{TPP}$:$C_{AIS}$ molar ratio (Figure 2d). The aggregation of TPP molecules is observed at a concentration of $C_{TPP}$ = 10$^{-6}$ M, corresponding to n = 2, where a decrease in FRET efficiency is also observed. This suggests that the TPP aggregation can be an additional energy relaxation process that occurs concurrently with FRET, resulting in a shortening of the FRET efficiency value.

**2.4 Spectrally Resolved FRET Dynamics in the CS-AIS-TPP Nanocomposites**

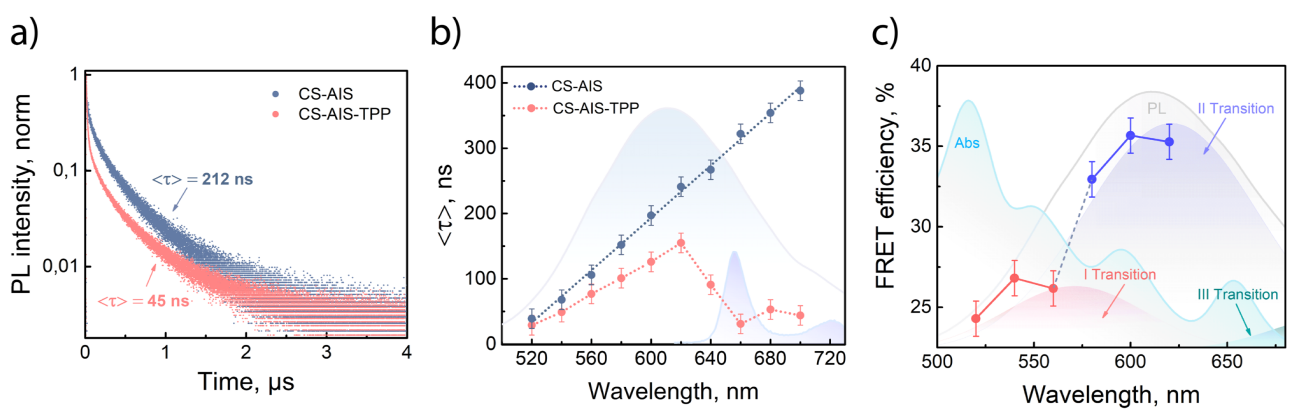

**Figure 3.** (a) The photoluminescence decay curves of CS-AIS and CS-AIS-TPP nanocomposites. (b) The dependence of the average photoluminescence lifetime on the detection wavelength. At the background the photoluminescence spectra of AIS QDs and TPP molecules are presented. The photoluminescence excitation wavelength is 420 nm. (c) The FRET efficiency calculated taking into account the photoluminescence lifetime of the donor in the presence and absence of the acceptor at different detection wavelengths. At the background the absorption spectrum of TPP molecules and the photoluminescence spectrum of AIS QDs with different radiative transitions are presented. The photoluminescence excitation wavelength is 420 nm.

**Table 1.** The fitting parameters for photoluminescence decay curves of the samples: photoluminescence lifetime τ, amplitude A.

| Sample | <τ>, ns | τ$_1$, ns | A$_1$, % | τ$_2$, ns | A$_2$, % | τ$_3$, ns | A$_3$, % | τ$_4$, ns | A$_4$, % |
|---|---|---|---|---|---|---|---|---|---|
| CS-AIS | 212 ± 20 | 658 ± 50 | 17 ± 2 | 200 ± 20 | 43 ± 5 | 33 ± 5 | 40 ± 5 | - | - |
| CS-AIS-TPP | 45 ± 5 | 617 ± 50 | 2 ± 1 | 190 ± 20 | 10 ± 1 | 44 ± 5 | 15 ± 2 | 8.6 ± 1 | 73 ± 5 |

It should be noted that the origin of the broad photoluminescence spectrum of ternary AIS QDs is not yet fully understood. Observations suggest that the emission of ternary QDs does not originate from exciton recombination as in the case of conventional binary nanocrystals. In contrast, their photoluminescence is commonly explained by either the radiative recombination of donor-acceptor (D-A) pairs,[65] the self-trapped exciton model,[66] or the recombination of a localized hole with a conduction band electron.[67] The detailed spectrally resolved analysis of the photoluminescence kinetics of the formed nanocomposites can provide insights into the mechanisms of non-radiative processes taking place within the system and lead to a better understanding of the recombination dynamics of AIS QDs.

We have analyzed the photoluminescence decay of CS-AIS-TPP nanocomposites with a molar ratio of the components of n = $C_{TPP}$:$C_{AIS}$ = 1:1, in which the TPP molecules remain mainly in the monomeric form, as mentioned above (Figure 2d). Since molecular aggregation can be accompanied by an increase of the non-radiative recombination rates, thus affecting the characteristic photoluminescence lifetime values, the choice of a sample with monomeric TPP is crucial for the correct FRET efficiency estimation. As shown in Figure 3a, the photoluminescence decay of the CS-AIS-TPP nanocomposites is faster than that of the CS-AIS solution. The photoluminescence decay curve of CS-AIS has been fitted by a three-exponential function according to Formula 2. In the case of CS-AIS-TPP nanocomposites, the appearance of the additional short component associated with TPP emission is observed, so the more accurate four-exponential fitting model has been applied. Analysis of the fitting parameters indicates the average photoluminescence lifetimes of (212 ± 20) ns and (45 ± 5) ns for CS-AIS and CS-AIS-TPP, respectively (Table 1). Such a decrease in the average photoluminescence lifetime of AIS QDs in the presence of TPP molecules can be attributed to the effective photoluminescence quenching of the long-lifetime QDs due to FRET to the short-lifetime TPP molecules. However, since the average photoluminescence lifetime is estimated as a weighted average in accordance with Formula 3, this value may be artificially lowered since the short-lived component of the TPP exhibits the largest amplitude contribution, as shown in Table 1.

To address this issue, we have performed the spectrally resolved study of the photoluminescence decay kinetics of the nanocomposites by using the bandpass filters to selectively detect the emission in a wavelength range corresponding to the photoluminescence spectrum of the QDs. Such an approach allows us to evaluate the FRET-associated quenching rate of QD photoluminescence in the spectral region of 520-640 nm, limiting the contribution of TPP emission. As shown in Figure 3b, the average photoluminescence lifetime of the CS-AIS solution increases linearly with increasing detection wavelength, which is a commonly observed dependence for AIS QDs. Ogawa et al. suggested that the characteristic behavior of the photoluminescence lifetime of ternary I-III-VI nanocrystals depending on the emission energy is associated with the D-A pair recombination.[65] Due to the distribution of the D-A pair distance, the recombination rate for a high-energy transition becomes higher compared to that for a low-energy transition, resulting in a shorter photoluminescence lifetime at higher emission energies. Further analysis reveals that the linear behavior of the dependence is maintained for the CS-AIS-TPP nanocomposites in the 520-640 nm spectral region, and the decrease of the slope is observed. The shortening of the donor QDs photoluminescence lifetime in the presence of the acceptor TPP molecules clearly indicates the formation of the additional non-radiative energy relaxation channel associated with FRET. Of particular note is the spectral dependence of the photoluminescence lifetime shortening of AIS QDs, which will be discussed in more detail below.

On the basis of the photoluminescence decay study results, we have estimated the FRET efficiency between AIS QDs and TPP molecules within the nanocomposite using the following formula:[62]

$$E_{FRET} = 1 - \frac{\tau_{DA}}{\tau_D}, \qquad (6)$$

where $\tau_{DA}$ and $\tau_D$ are the photoluminescence lifetimes of the donor in the presence and absence of an acceptor, respectively. Evaluation of the spectrally resolved dynamics of FRET reveals that the efficiency of non-radiative energy transfer varies for different emission energies of AIS QDs, as shown in Figure 3c. To explain these observations, we propose a model based on the existence of multiple radiative D-A transitions in AIS QDs. This aspect is consistent with the description proposed by Mao et al. who suggest that the photoluminescence of AIS QDs arises from the intrinsic trap states of different energies formed between interstitial atoms or vacancies which are responsible for D-A transitions.[68] In this context, the emission of AIS QDs has been deconvoluted by three Gaussian profiles (Figure S4 and Table S3), which are ascribed to the multiple radiative transitions (Figure 3c). We suggest that the excitation energy can be non-radiatively transferred from each of the radiative transitions of the AIS QD donors to TPP acceptors, resulting in the variation of the FRET efficiency for different energy states. As can be seen in Figure 3c, the average FRET efficiency associated with the I Transition of AIS QDs equals approximately 27%, while for the II Transition this value exceeds 35%. Such a result is consistent with the statement that the FRET rate depends strongly on the degree of spectral overlap (Figure S5). Notably, the obtained values are close to the $E_{FRET}$ = 32.8% obtained from the FRET efficiency calculations using the TPP sensitization rate (Section 2.2).

## 2.5 FRET-Enhanced Singlet Oxygen Generation by the CS-AIS-TPP Nanocomposites

It has been previously discovered that the FRET-driven excitation of the porphyrin molecules can provide a higher efficiency of SO generation compared to the direct excitation of the molecules at the same wavelength due to the high extinction coefficients of QD energy donors[43]. To experimentally validate the enhanced SO generation capability of CS-AIS-TPP nanocomposites, the Singlet Oxygen Sensor Green (SOSG) probe, which is selective for detecting SO[69], has been employed. The SOSG has been mixed with the reference CS-TPP solution and CS-AIS-TPP nanocomposites and the samples have been irradiated with a 460 nm diode source. The results demonstrate that the photoluminescence of SOSG exhibits a more rapid increase in the presence of CS-AIS-TPP nanocomposites under further irradiation, compared to the CS-TPP solution (Figure S6). This observation indicates that the formed nanocomposites exhibit a more pronounced generation of SO.

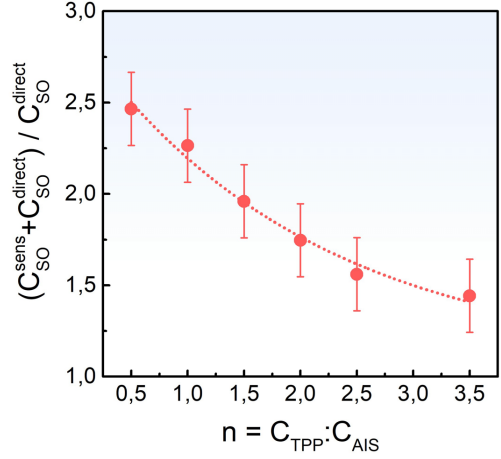

**Figure 4.** The dependence of the relative SO concentration enhancement of TPP within the CS-AIS-TPP nanocomposites on the n = $C_{TPP}$:$C_{AIS}$ molar ratio.

Furthermore, we propose a model to theoretically evaluate the relative enhancement of the SO generation by TPP molecules through FRET from AIS QDs within the formed CS-AIS-TPP nanocomposites. The concentration of the SO produced by the CS-incorporated TPP molecules under the external irradiation can be estimated as follows:

$$C_{SO}^{direct} \sim C_{TPP} \cdot \varepsilon_{TPP} \cdot m \cdot E_{SO}^{TPP}, \tag{7}$$

where $C_{TPP}$ is the concentration of the TPP molecules; $\varepsilon_{TPP}$ is an extinction coefficient of the TPP molecules; $m$ is the fraction of the TPP molecules in monomeric form; $E_{SO}^{TPP}$ is the efficiency of the SO generation by TPP molecules. The concentration of the FRET-enhanced SO produced by the TPP molecules within the CS-AIS-TPP nanocomposite under the external irradiation is given with:

$$C_{SO}^{sens} \sim E_{FRET} \cdot C_{AIS} \cdot \varepsilon_{AIS} \cdot E_{SO}^{TPP} \cdot F, \tag{8}$$

where $E_{FRET}$ is the FRET efficiency between AIS QDs and TPP molecules within the nanocomposites; $C_{AIS}$ is the concentration of AIS QDs; $\varepsilon_{AIS}$ is an extinction coefficient of the AIS QDs; $F$ is the photoluminescence quenching efficiency of the AIS QDs. Therefore, the relative enhancement of the SO generation by TPP molecules via FRET from AIS QDs can be calculated according to the formula:

$$\frac{C_{SO}^{sens} + C_{SO}^{direct}}{C_{SO}^{direct}} = \frac{E_{FRET} \cdot C_{AIS} \cdot \varepsilon_{AIS} \cdot F}{C_{TPP} \cdot \varepsilon_{TPP} \cdot m} + 1 \tag{9}$$

The results of the above calculations are illustrated in Figure 4, clearly indicating the enhancement of the SO generation within the CS-AIS-TPP nanocomposite for the whole range of n = $C_{TPP}$:$C_{AIS}$ molar ratio. Specifically, the maximum value of the increase in SO concentration is determined to be 2.5 at a component molar ratio of n = 0.5. As the TPP concentration within the CS-AIS-TPP nanocomposite increases, the gradual decrease of the SO generation enhancement can be observed, indicating the presence of TPP aggregates which are not capable of producing SO, but can still receive energy from AIS QDs.

## 3 Conclusions

In summary, we have proposed an approach for fabricating 150 nm in size functional nanocomposites based on hydrophobic TPP molecules and hydrophilic ternary AIS QDs incorporated into a CS matrix. We have estimated that TPP molecules are present completely in a monomeric form within the nanocomposite in a wide range of concentrations. The detailed time- and spectrally resolved analysis of the FRET dynamics between QDs and TPP molecules confirms the concept of the presence of multiple radiative transitions in ternary AIS QDs. Moreover, the variation of FRET efficiency within the nanocomposite for different energy states of QDs has been observed. Based on these results, we further propose the mechanism of SO efficiency enhancement via FRET. We believe that our study provides new insights into the design of FRET-amplified photosensitizers based on non-toxic ternary I–III–VI QDs for a-PDT and leads to a better understanding of the recombination dynamics of AIS QDs. In future studies, we aim to expand the scope of our work and conduct a comprehensive investigation into the a-PDT effect of the formed nanocomposites on various types of Gram-positive and Gram-negative bacterial cells.

# 4 Experimental Section

### 4.1 Chemicals

Indium (III) chloride, silver nitrate, zinc (II) acetate dihydrate, $Na_2S \cdot 9H_2O$, $NH_4OH$ (aqueous 5 M solution), mercaptoacetic acid (MAA), 2-propanol, chitosan (50–190 kDa), acetic acid, and tetrachloromethane were purchased from Sigma-Aldrich (USA). TPP and SOSG chemical probe were purchased from Frontier Scientific (USA). All chemicals were used without further purification.

### 4.2 Synthesis of AgInS$_2$/ZnS Quantum Dots

Colloidal core-shell AgInS$_2$/ZnS (AIS) QDs were synthesized according to a previously published procedure.[70] Briefly, 1 mL of aqueous 0.1 M AgNO$_3$ solution, 2 mL of aqueous 1 M MAA solution, and 0.2 mL of aqueous 5 M NH$_4$OH were added to 96 mL water under magnetic stirring. The resulting suspension became colorless after the addition of 0.45 mL of aqueous 5 M NH$_4$OH solution and 0.7 mL of aqueous 1 M InCl$_3$ solution containing 0.2 M HNO$_3$. Then, 1 mL aqueous 1 M Na$_2$S solution was added at stirring and the solution was heated in a water bath at 90–95°C for 30 min. Next, 0.5 mL of aqueous 1 M MAA was added again and the solution evaporated at 40°C. For the ZnS shell growth, 1 mL of aqueous 1 M MAA solution and 1 mL of aqueous 1 M Zn(CH$_3$COO)$_2$ solution were added under intensive magnetic stirring and the resulting solution was additionally heated for 30 min. As a result, MAA-capped AIS QDs with an average diameter of 2.7 nm were obtained (Figure S7).

### 4.3 Nanocomposites Formation

For nanocomposites formation, CS (4 mg/ml) was dissolved in 0.1% acetic acid aqueous solution under magnetic stirring for 24 hours. After that, the CS solution was filtered using 0.22 µm pore size membrane filter. Next, an aqueous solution of MAA-stabilized AIS QDs ($C_{AIS}$ = 10$^{-6}$ M) was added to the CS solution in a volume ratio of $V_{CS}:V_{AIS}$ = 1:1, and the mixture was left on a magnetic stirrer for 2 hours. Then, portions of TPP molecules ($C_{TPP}$ = 6.7 · 10$^{-4}$ M) in CCl$_4$ were added to the solution under intensive magnetic stirring. The time intervals for adding portions of TPP were chosen to achieve the complete evaporation of the organic solvent. As a result, a series of CS-AIS-TPP nanocomposite solutions with different molar ratios of QDs and TPP molecules were obtained. In addition, one-component CS-TPP and CS-AIS solutions were prepared for the reference.

### 4.4 Characterization Methods

High resolution transmission electron microscopy (HR-TEM) images were recorded with a FEI Titan G3 transmission electron microscope. For QD diameter estimation at least 80 particles were analyzed with ImageJ software. The morphology and sizes of the nanocomposites were analyzed using the modified Stripe Fluorescence Recovery After Photobleaching (s-FRAP) technique implemented on the basis of a laser scanning microscope LSM 710 (Carl Zeiss, Germany). The diode laser source with a wavelength of 405 nm focused with 20x/0.4 and 20/0.75 lenses, was used both for the exposure and the photoluminescence excitation, with the respective laser power ratio 100:1. Dynamic Light Scattering (DLS) and zeta-potential measurements were performed using the Zetasizer Nano ZS instrument (Malvern Panalytical, UK).

### 4.5 Optical Measurements

The absorption spectra were recorded using the UV-3600 spectrophotometer (Shimadzu, Japan) equipped with an integrating sphere. The photoluminescence spectra of the samples were measured using the Cary Eclipse fluorescence spectrophotometer (Varian, Australia). The photoluminescence quantum yield (QY) of QDs was estimated by a comparative method using Rhodamine 6G with standard QY of 95% as a reference fluorophore:

$$QY_{QD} = QY_R \cdot \frac{I_R}{I_{QD}} \cdot \frac{D_{QD}}{D_R} \cdot \frac{n_R^2}{n_{QD}^2}, \qquad (10)$$

where $I$ is the integrated photoluminescence intensity, $D$ is the optical density on the photoluminescence excitation wavelength, and $n$ is the reflective index of a solvent. The subscript $R$ stood for the reference.

The photoluminescence decay kinetics was studied using a MicroTime100 time-correlated single photon counting fluorescence microscope (PicoQuant, Germany). A pulsed laser with a wavelength of 409 nm operating at 250 kHz was used for photoluminescence excitation. Bandpass filters with a full-width at half maximum of 10 nm were employed to selectively detect the emission in a wavelength range corresponding to the photoluminescence spectra of the samples. The photoluminescence decay curves were fitted by a function:

$$I(t) = I_0 + \sum_i A_i \exp\left(-\frac{t}{\tau_i}\right), \tag{11}$$

where $A_i$ and $\tau_i$ are the amplitude and the characteristic lifetime of the *i*-th component, respectively. The average photoluminescence lifetime was defined as:

$$\langle \tau \rangle = \frac{\sum_i A_i \tau_i^2}{\sum_i A_i \tau_i} \tag{12}$$

**4.6 SO Generation Study**

To assess the ability of the nanocomposites to generate singlet oxygen (SO), we utilized the Singlet Oxygen Sensor Green (SOSG) chemical probe. The samples were prepared by mixing the nanocomposite solutions with a 5 mM solution of SOSG. Subsequently, the samples were subjected to irradiation from a 460 nm diode source with a power density of 3 mW/cm$^2$. The photoluminescence spectra of SOSG in the presence of the nanocomposites were measured at 530 nm, with excitation at 500 nm, at regular intervals.

**Author Contributions**

T. O. O.: conceptualization, methodology, investigation, validation, visualization, writing – original draft; A. A. M.: methodology, investigation; L. N. B.: methodology, investigation; E. S. S.: methodology, investigation; A. I. D.: methodology, investigation; F. A. S.: methodology, A. V. V.: methodology; E. O. M.: methodology; A. O. O.: funding acquisition, validation, supervision, writing – review & editing.

**Conflicts of interest**

The authors declare no competing financial interest.

**Acknowledgements**

This work was financially supported by the Ministry of Education and Science of the Russian Federation, State assignment, Passport 2019-1080 (Goszadanie 2019-1080) and by Clover Program: Joint Research Projects of Skoltech, MIPT, and ITMO. Tatiana O. Oskolkova and Lyubov' N. Borodina acknowledge the support by RPMA grant of the School of Physics and Engineering of ITMO University for students. The AICF of Skoltech acknowledges for providing access to the TEM facilities and Maria Kirsanova for microscope operation.

# Supporting Information

**S1. Size and Morphology Characterization of the CS-AIS-TPP Nanocomposites**

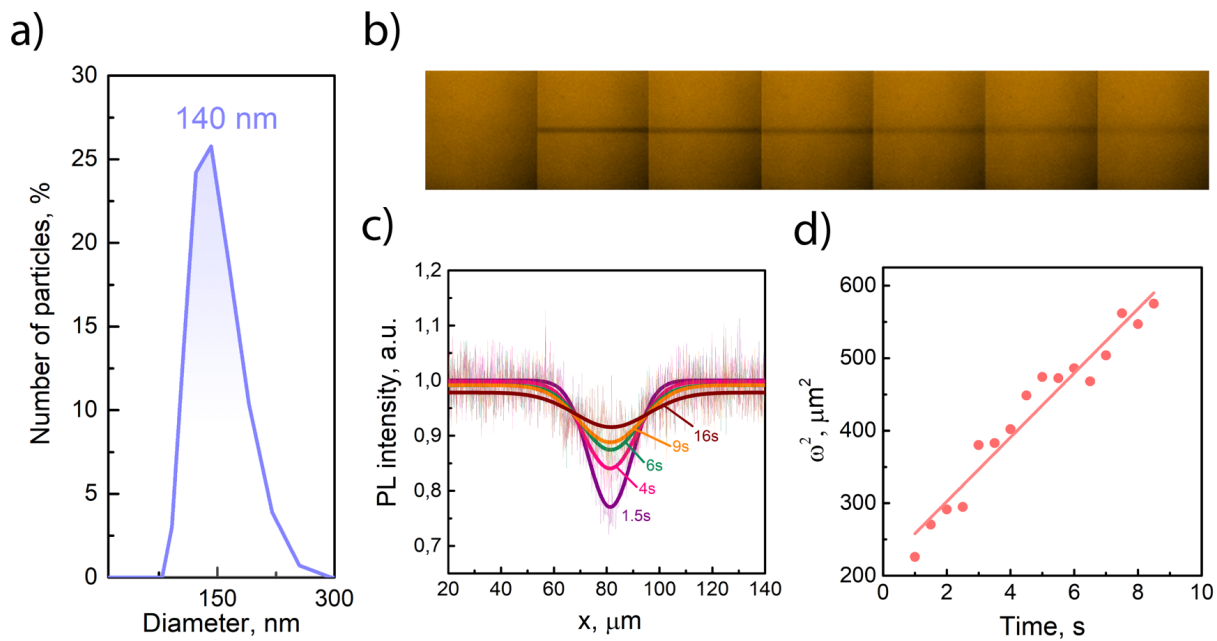

**Figure S1.** (a) The size distribution of the CS-AIS-TPP nanocomposites obtained using the Dynamic Light Scattering (DLS) technique. (b) A sequence of the photoluminescence images of a 153x153 µm section of a CS-AIS-TPP nanocomposites colloidal solution at a layer thickness of 120 µm taken using a LSM 710 laser scanning microscope under excitation by 405 nm laser light before (left, 0 s) and every 0.48 s after exposing a 20 µm wide band with the same irradiation. (c) Transverse luminescence intensity profiles in the vicinity of the photobleached band of the CS-AIS-TPP nanocomposite solution after exposure (1.5 s) and 4 s, 6 s, 9 s, 16 s after it, obtained by averaging the luminescence intensity values of images over 1000 columns along the band (noisy thin lines) and the results of their Gaussian approximation (broad smooth curves). (d) Time dependence of the square of the Gaussian width parameter of the photobleached band profile in the colloidal solution of CS-AIS-TPP nanocomposites and the result of its linear approximation.

The technique, referred to as Stripe-FRAP (s-FRAP)[1], involves the following steps: (i) exposure of a narrow stripe in a colloidal solution of luminescent particles with laser light, which leads to the change in their photoluminescence quantum yield, hence photoluminescence intensity; (ii) tracking the changes in photoluminescence pattern consisting in smearing-out of the exposed stripe caused by particle diffusion; (iii) the analysis of as a linear temporal dependence of the squared width of the photoluminescence transversal profile, which slope is proportional to the diffusion coefficient, and respectively inversely proportional to the particle radius, according to the Stokes-Einstein relation (Formula 1):

$$D = \frac{k_B T}{6\pi\eta}, \qquad (1)$$

where $D$ is a diffusion coefficient, $k_B$ is the Boltzmann constant, $T$ is the absolute temperature, $\eta$ is the macroscopic dynamic viscosity, and $r$ is the hydrodynamic radius. To elucidate the technique, we present a sequence of photoluminescence images (Figure S1b), a corresponding series of transversal profiles (Figure S1c), and their squared widths (Figure S1d).

## S2. The FRET Efficiency Calculations

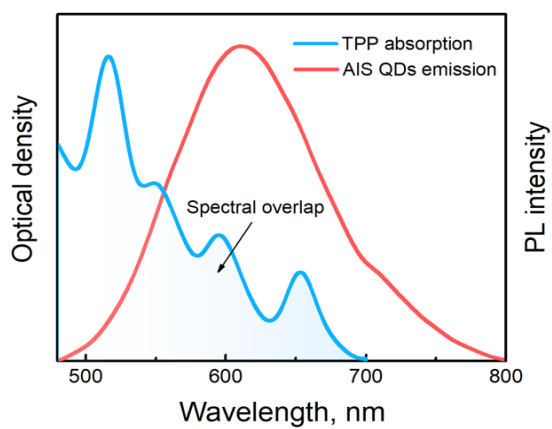

**Figure S2.** The absorption spectrum of the acceptor tetraphenylporphyrin (TPP) molecules and the photoluminescence spectrum of the donor $AgInS_2/ZnS$ (AIS) quantum dots (QDs). The photoluminescence excitation wavelength is 420 nm.

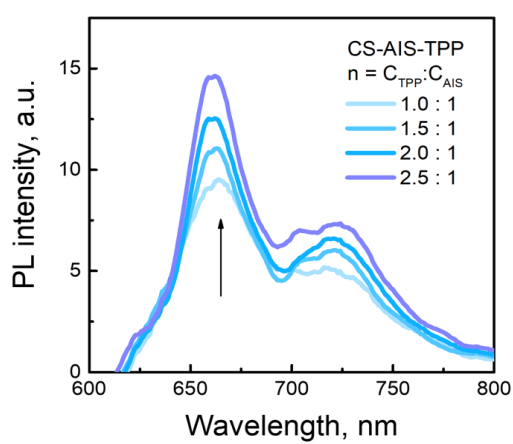

**Figure S3.** The sensitized photoluminescence spectra of the TPP molecules within the CS-AIS-TPP nanocomposites with different $n = C_{TPP}:C_{AIS}$ molar ratios. The photoluminescence excitation wavelength is 460 nm.

**Table S1.** Parameters for the calculations of the Förster resonance energy transfer (FRET) efficiency ($E_{FRET}$) within the CS-AIS-TPP nanocomposites with different $n = C_{TPP}:C_{AIS}$ molar ratios: the optical density D, the photoluminescence quenching efficiency of the donor F, the photoluminescence intensity I.

| $n = C_{TPP}:C_{AIS}$ | $D_{460\ nm}$ | $D_{600\ nm}$ | F | $I_{460\ nm}$ | $I_{600\ nm}$ | $E_{FRET}$, % |
|---|---|---|---|---|---|---|
| 0.5 : 1 | 0,12 | 0,01 | 0,30 | 5,97 | 5,71 | 30,5 ± 3 |
| 1.0 : 1 | 0,15 | 0,02 | 0,52 | 9,69 | 7,01 | 32,8 ± 3 |
| 1.5 : 1 | 0,17 | 0,03 | 0,51 | 10,95 | 9,48 | 33,8 ± 3 |
| 2.0 : 1 | 0,19 | 0,03 | 0,50 | 12,64 | 11,59 | 33,0 ± 3 |
| 2.5 : 1 | 0,21 | 0,03 | 0,50 | 14,11 | 15,67 | 29,1 ± 3 |
| 3.5 : 1 | 0,25 | 0,05 | 0,55 | 14,02 | 17,89 | 27,8 ± 3 |

**Table S2.** Parameters for the calculations of the FRET efficiency ($E_{FRET}$) within the CS-AIS-TPP nanocomposites with different $n = C_{TPP}:C_{AIS}$ molar ratios: the Förster distance $R_0$, the overlap integral between the donor emission band and the acceptor absorption band $I_{overlap}$, the distance between the donor and acceptor R.

| $n = C_{TPP}:C_{AIS}$ | $R_0$, nm | $I_{overlap}$ | $E_{FRET}$, % | R, nm |
|---|---|---|---|---|
| 0.5 : 1 | 8.2 ± 1.0 | $4.0 \cdot 10^{-12}$ | 30.5 ± 3 | 9.4 ± 1.0 |
| 1.0 : 1 | 8.3 ± 1.0 | $4.4 \cdot 10^{-12}$ | 32.8 ± 3 | 9.4 ± 1.0 |
| 1.5 : 1 | 8.3 ± 1.0 | $4.4 \cdot 10^{-12}$ | 33.7 ± 3 | 9.3 ± 1.0 |
| 2.0 : 1 | 8.3 ± 1.0 | $4.3 \cdot 10^{-12}$ | 33.0 ± 3 | 9.3 ± 1.0 |

## S3. The Recombination Dynamics of AIS QDs

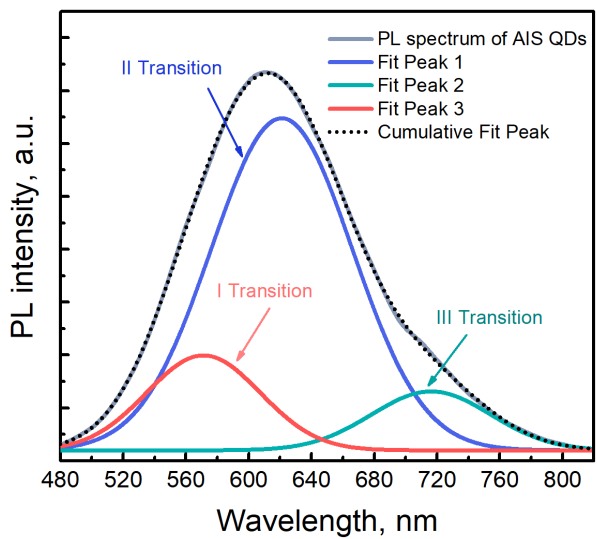

**Figure S4**. Gaussian deconvolution of the photoluminescence spectrum of AIS QDs in a CS matrix.

**Table S3.** Parameters for the Gaussian deconvolution of the photoluminescence spectrum of AIS QDs in a CS matrix.

|  | Peak position, nm | Adjusted $R^2$ |
|---|---|---|
| Fit Peak 1 | 571.13 | 0.99984 |
| Fit Peak 2 | 621.04 |  |
| Fit Peak 3 | 716.22 |  |
| Cumulative Fit Peak | 611.38 |  |

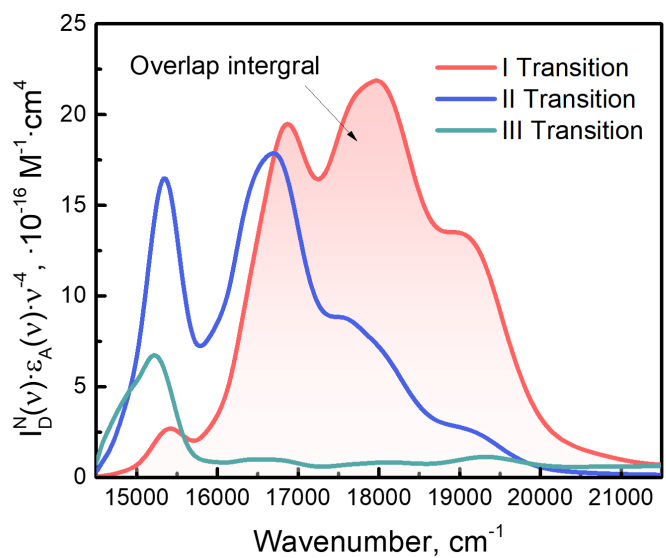

**Figure S5.** The overlap integral between the acceptor TPP absorption band and different radiative transitions of the donor AIS QDs.

## S3. The Singlet Oxygen Generation Study

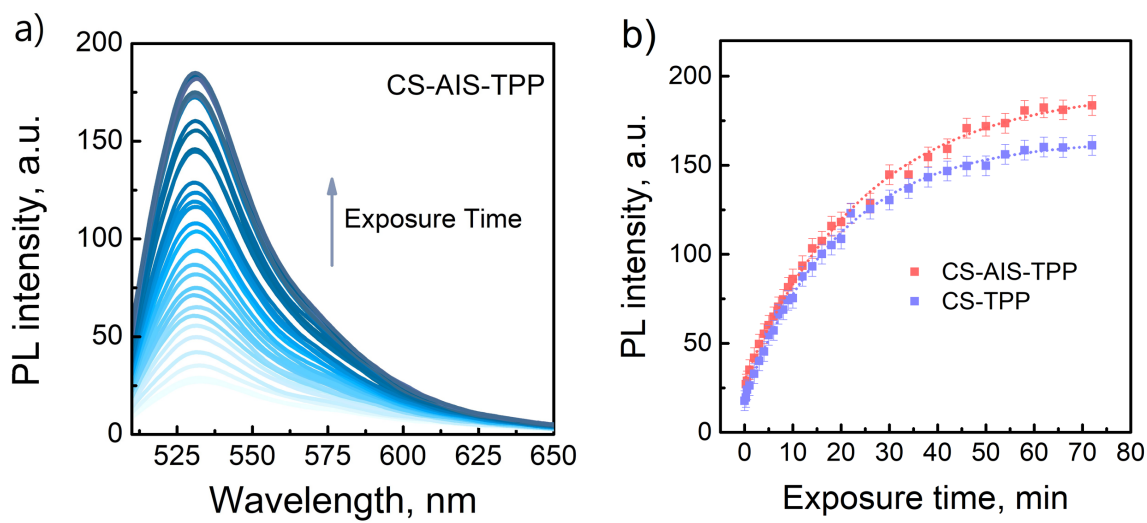

**Figure S6.** (a) The photoluminescence spectra of SOSG in the presence of CS-AIS-TPP nanocomposites under the 460 nm light irradiation. (b) The photoluminescence intensity of SOSG in the presence of CS-TPP and CS-AIS-TPP nanocomposites at 530 nm as a function of the exposure time.

## S4. Size Characterization of the AIS QDs

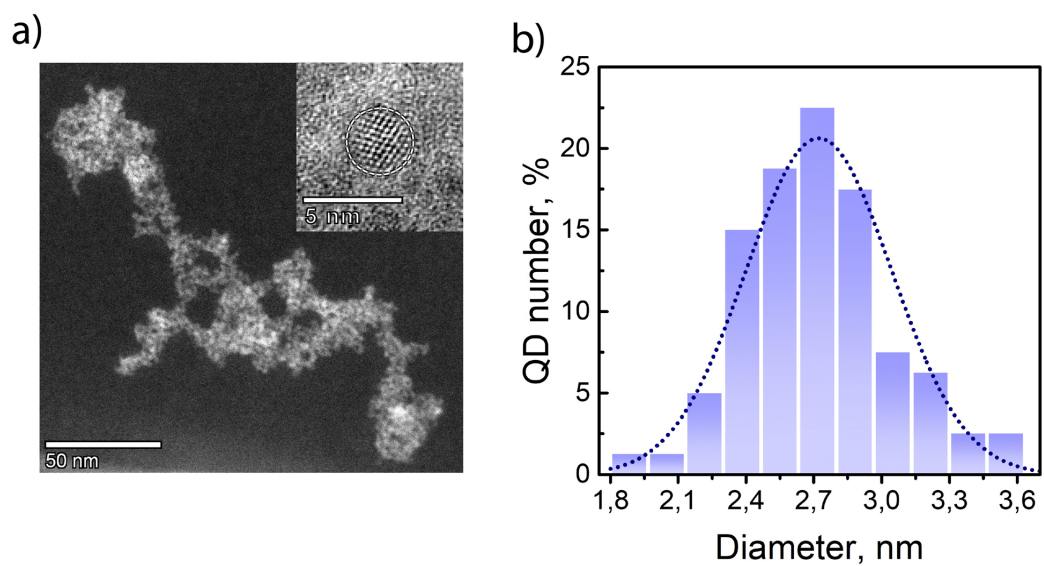

**Figure S7.** (a) The STEM image of MAA-caped AIS QDs. Inset shows the HTEM image of a single MAA-caped AIS QD. (b) The size distribution histogram of MAA-caped AIS QDs.